\documentclass{aastex}          %% The manuscript based on AASTeX v5.x
\usepackage{spr-astr-addons}    %% mimicing ASTR journal style
\begin{document}
\title{UV Properties of Type Ia Supernova and their Host Galaxies}

\shorttitle{UV SN Ia}
\shortauthors{B. E. Tucker}

\author{Brad. E. Tucker\altaffilmark{1}, The ESSENCE Project}

\altaffiltext{1} {The Research School of Astronomy and Astrophysics, Australian National University, Mount Stromlo Observatory, via Cotter Road, Weston Creek, ACT 2611, Australia; brad@mso.anu.edu.au}

\begin{abstract}
Type Ia Supernova(SN Ia) are a powerful, albeit not completely understood, tool for cosmology.
Gaps in our understanding of their progenitors and detailed physics can lead to systematic errors in the cosmological distances they measure.  We use UV data in two context to help further our
understanding of SN Ia progenitors and physics.  We analyze a set of nearly 700 light curves, and find no signature of the shock heating of a red giant companion, predicted by Kasen (2010), casting doubt as to frequency of this SN Ia channel.  We also use UV imaging of high redshift host galaxies of SN Ia to better understand the environments which SN Ia occur.  We show that some high-z elliptical galaxies have current star formation, hindering efforts to use them as low-extinction environments.  We show cosmological scatter of SN distances at large effective radii in their hosts is significantly reduced, and argue this is due to the smaller amounts of dust affecting the SN Ia.  Finally, we find a two component dependance of SN distance measurements as a function of their host galaxy's FUV-V color.  This indicates that both the age and metallicity/mass of the host galaxy maybe important ingredients in measuring SN Ia distances. 
\end{abstract}

\section{Introduction}

\ The use of type Ia Supernova (SN Ia) in cosmology is a long and arduous road.  The empirical relation of the light curve shape of a SN Ia to its peak absolute magnitude has become a cornerstone of measuring the cosmic acceleration of the Universe \citep{Phil93, Hamuy96b, RPK96, Goldhaber01, Prieto05, Guy07, Jha07, Conley08}.  However we still lack a clear understanding of the progenitor system \citep{HN00} and their use to understand the nature of Dark Energy is becoming limited by systematics \citep{Astier06,MWV07} rather than statistics.

\  It is widely held that the progenitor of a SN Ia is a carbon/oxygen white dwarf (WD) which accretes mass in a binary process until a thermonuclear runaway begins - typically as the white dwarf's mass approaches the $1.4M_{\odot}$ Chandrasekhar limit.  In contrast to core collapse SN, pre-explosion imaging has not turned up explosion progenitors, confirming the relatively faint nature of SN Ia progenitor systems \citep{MaozMannucci08}.  The mostly widely held model, the single degenerate (SD) scenario, has a non-degenerate companion, such as a red giant (RG) or main sequence (MS) star depositing material onto the WD.  This companion star should remain relatively bright and potentially with signatures of rotation and spatial motion after the explosion. \citep{Wheeler75,FryxellArnett81,Livne92}.   \citet{Pilar04,Pilar09} have identified a potential companion candidate in the remnant of SN 1572 which has a high spatial velocity and some other signatures expected for a SN Ia companion star.  \citet{WEK09} have investigated further the potential candidate and other stars in SN1572, and while unable to rule out the candidate entirely, they argue it is an unlikely SN Ia companion star. 

\ A second possible scenario is the double degenerate (DD) scenario,  in which two WDs coalesce, exceeding the Chandrasekhar limit \citep{IT84,Webb84}. \citet{Howell06} and \citet{Hick07} believe to have observational evidence of a SN Ia exploding due to the DD scenario, based on a few exceptional objects that appear to have synthesized more than a Chandrasekhar Mass of $^{56}$Ni, and are estimated to have ejected approximately two solar masses of material.  These claims have some model dependence, but seem compelling.  Despite these instances, such SN are unusual and do not represent the average population of SN Ia's.

\  Investigating the environments of SN Ia allows an alternative way to examine SN Ia properties, while providing a potential avenue to help alleviate systematic problems in their application as distance indicators.  Early studies suggested that SN Ia are more frequently associated with younger stellar populations in star-forming late type galaxies \citep{OT79},  and prefer the disks of the galaxies rather than bulges \citep{WHW97}.  However, \citet{MC96} observed that SN Ia's occur far from the spiral arms of late-type galaxies, and in galaxies with low star formation rates (SFR).  These observations argue for a delay between the formation of the systems and the subsequent explosion as a SN.  \citet{Hamuy96b}  found a loose correlation between the SN Ia decline rate and the host galaxy type.  They found the faint, fast declining SN occur in early-type galaxies, whereas the bright, slower declining SN Ia occur exclusively in late-type star-forming galaxies.  This trend has been confirmed with other studies \citep{Neill09} and at higher redshifts \citep{Sullivan06}.  \citet{JG05} found that fast, faint declining SN Ia's were only in galaxies without any significant star-formation activity.  However, a follow up study \citep{JG08} found that those elliptical galaxies actually had some star-formation, contrary to previous thoughts.  

Studies using nearby SN Ia's \citep{Kelly10}, SN Ia's in the $0.1 < z < 0.3$ range \citep{Lampeitl10} and at higher redshifts ($0.3 < z < 1.0$) \citep{Sullivan10} have all found that the luminosity of SN Ia, corrected for light curve shape, depends on the host galaxy stellar mass (which correlates metallicity).  While the trends are weak, fitting for the trend can improve SN Ia distances by $\approx 5\%$ \citep{Sullivan10}.  

\section{Progenitor Signatures in UV Light Curves}

\ Recent developments in the observations of core-collapse SN have opened up a new means to understanding the progenitor systems of SN.  After the explosion of a core-collapse supernova, the shock wave propagates through the envelope of the star, emitting first in x-rays then the ultraviolet and optical.  This early shockwave has now been observed  in the X-ray with the accidental discovery of SN 2008D by SWIFT \citep{Soderberg08,Modjaz09}.  \citet[hereafter K10 and references therein]{Kasen10} has calculated a model applying similar physics to the companion stars of SN Ia.

\subsection{The Model}
\begin{figure}[htb]
\plotone{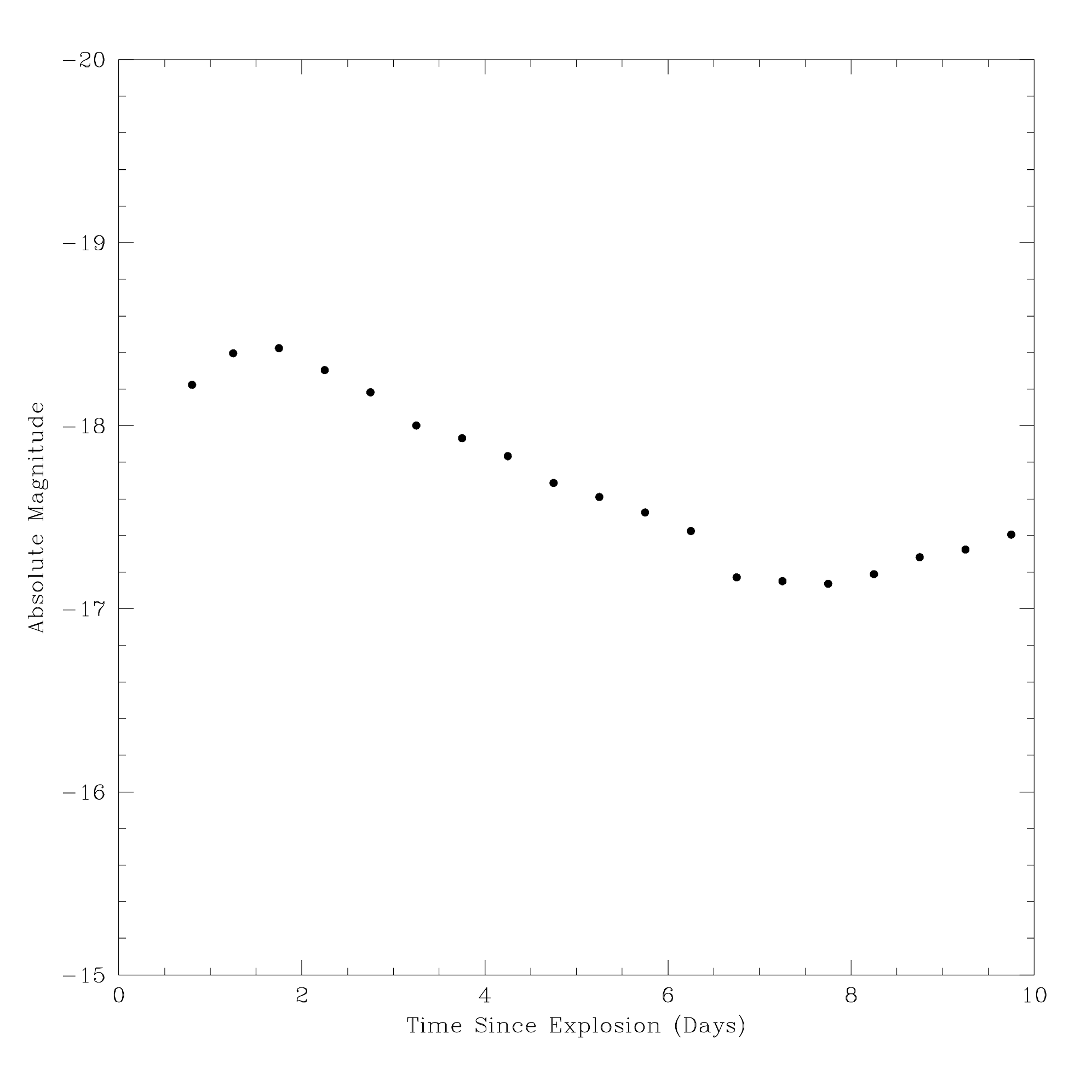}
\caption{Calculated model from K10 for a viewing angle of $ \Theta = 0 \deg$ in the $U$ filter.  We can see the shocked emission's signature in SN Ia light curves.  The RG companion has the biggest effect and should be easily discernible in observed SN Ia light curves.}
\label{fig:KasenLC}
\end{figure}

\ The model in K10 uses plausible non-degenerate companions:  $1-3 M_{\odot}$ and $5-6 M_{\odot}$ main sequence (MS) sub-giants and $1-2 M_{\odot}$ red giants(RG).  Upon explosion of the WD, the ejecta from the SN Ia will expand freely until hitting the companion.  As the ejecta hits the companion, a bow shock is formed, heating and compressing the remaining ejecta into a thin shell.  The companion subsequently diverts the incoming flow of material from the SN, creating a hole in the ejecta.  This hole allows the thin shell of heated ejecta to rapidly escape.  After a further time, the remaining ejecta fills in the hole created by the companion.

\ When the companion carves out a hole in the ejecta, the shocked ejecta radiates, and this radiation can escape through the cavity in the ejecta.  Both the radius of the shocked stellar material and the ejecta cavities are large in the case of RG companions, allowing a significant fraction of the radiation to escape, producing an observable signature on a SN Ia light curve. The effects for MS companions are significantly less. 

\ The emission from the prompt burst is mostly in soft X-rays, however energy not radiated in the prompt burst will be emitted in the following hours and days in the UV and optical. The strength of the observable emission will depend on the viewing angle of the observer, with the strongest emission seen with a viewing angle looking directly at the hole.  K10 calculates about $10 \%$ of all SN Ia have a favorable viewing angle with strong detectable emission in the ultraviolet and $B$ optical bands.  Figure \ref{fig:KasenLC} shows the strength of the shocked escaped emission at the ideal viewing angle, $ \theta = 0 \deg$, in the ultraviolet for the RG scenario.  

\subsection{Observations}

\ To test this prediction, we compiled 695 SN Ia light curves from several surveys \citep{UVExcess} over the redshift range of $0.03 < z < 1.0$.  The data consists of the Center for Astrophysics (CfA)1 \citep{CfA1}, CfA2 \citep{CfA2}, and CfA3 \citep{CfA3} data sets, the Sloan Digital Sky Survey-II (SDSS-II) \citep{Frieman08} first year data release \citep{Kessler09},  the Equation of State SuperNova trace Cosmic Expansion (ESSENCE) \citep{Gajus07,MWV07,GSN10} and the Supernova Legacy Survey (SNLS) \citep{Astier06}.  To be able to compare all of the data sets, we must correct them to a common frame.   This requires us to correct for extinction due to dust (both in the host galaxy and the Milky Way), time dilation, different filter systems and redshift filter corrections (k-corrections).  We transform all of the light curves into rest-frame $UBVRI$ at $z=0$ using the SALT2 package \citep{Guy07}.  With the transformations, we can compare it to the predictions of K10 and to the "normal" SN Ia template, the Leibundgut template \citep{Leibundgut89}

\subsection{Discussion}
\begin{figure}[htb]
\plotone{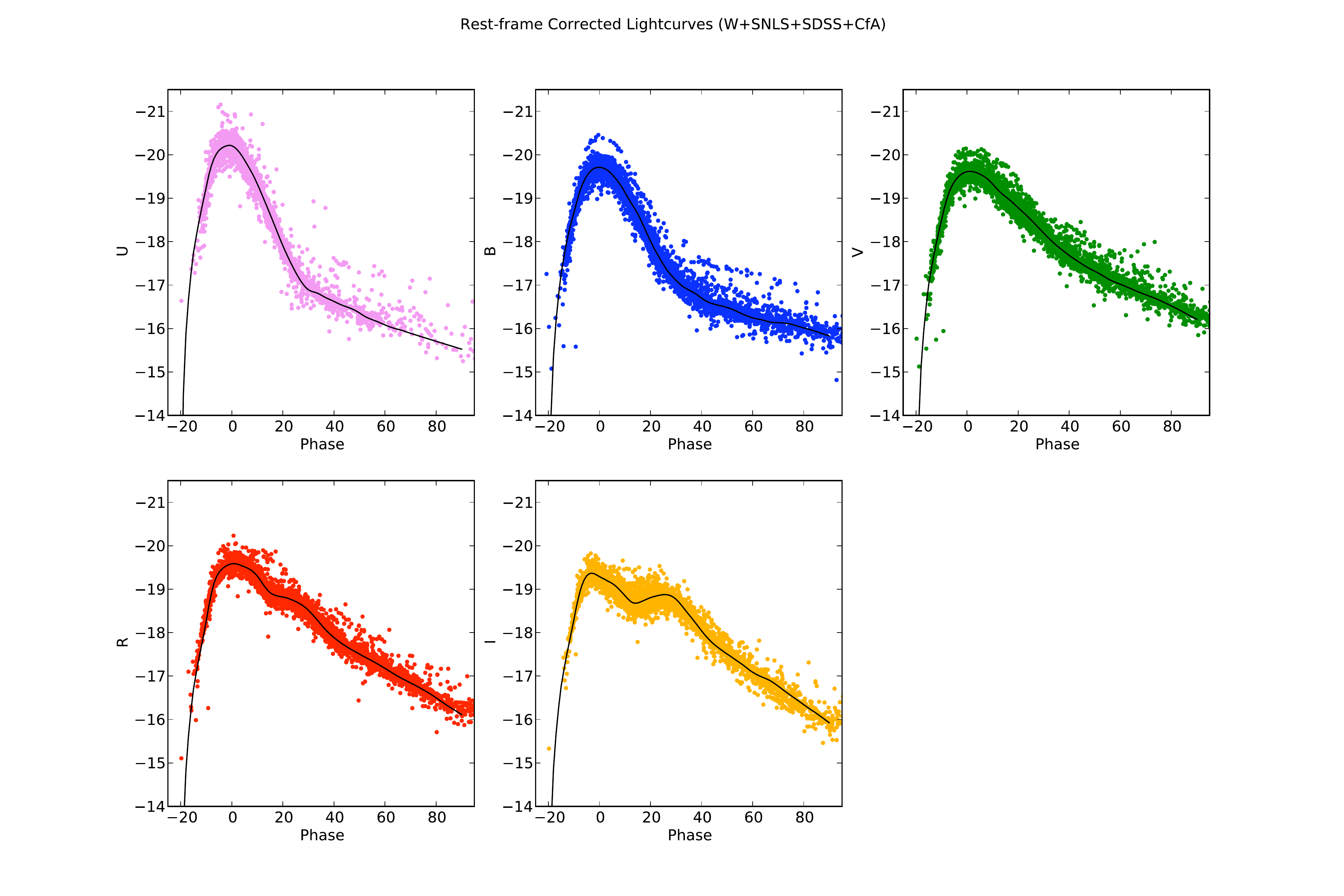}
\caption{A compilation of 695 light curves from the CfA1,2, and 3 surveys, SDSS-II, ESSENCE, and SNLS supernova surveys.  The light curves have been run through SALT2 in order to bring them to a common system, $UBVRI$.  The black line is the Leibundgut template of a standard SN Ia.  It is evident there is no detectable signature from the emission due to the shocked ejecta escaping a hole created by the companion.  There should be $\approx 70$ light curves that have this feature, with the RG scenario being the strongest contributor.  Despite the decrease in sampling at early times (where the feature occurs), no hint of the presence is seen, ruling out the RG scenario.}
\label{fig:RestFrameLC}
\end{figure}

\ Figure \ref{fig:RestFrameLC} shows our complied light curves from all of the surveys in rest-frame $UBVRI$, with the black line showing, as reference the Leibundgut template of a normal SN Ia.  For our purposes, we treat the $U$ filter as being approximated by the near-ultraviolet filter calculation as calculated in K10, with the $B$ and $V$ filters being direct comparisons to K10's model predictions.  Light curve variations are larger after maximum than before, because variations in the amount and location of energy ejection caused by the decay of $^{56}$Ni are most pronounced after maximum light.  \citep{KasenWoosley07,Hayden10}.  While global fits to $U$ and $B$ bands show some dispersion due to intrinsic variation and the effects of dust, these departures are much smaller than the signal predicted by K10.

\ From K10, we expect, $10 \%$ of all light curves should show the signature of the companion star.  If all SN Ia have RG companions, our data set should have $\approx$ 70 light curves with this feature.  At $t \lesssim 7$ days, where the predicted effects are most profound, we should be able to see the signature of the companion, with the signature increasing as $t \Rightarrow 0$.  Examining our light curves in $UBV$, we detect no signature in any filter for any event.  While our sampling greatly decreases in early time measurements, we have a large enough sample to yield a few detections if a majority of SN Ia have progenitors involve RG companions.   Our data set observations are well fit by the Leibundgut template, showing the analysis is consistent.  \citet{UVExcess} will publish formal limits on the fraction of SN Ia with Red Giant companions.

\begin{figure}[htb]
\plotone{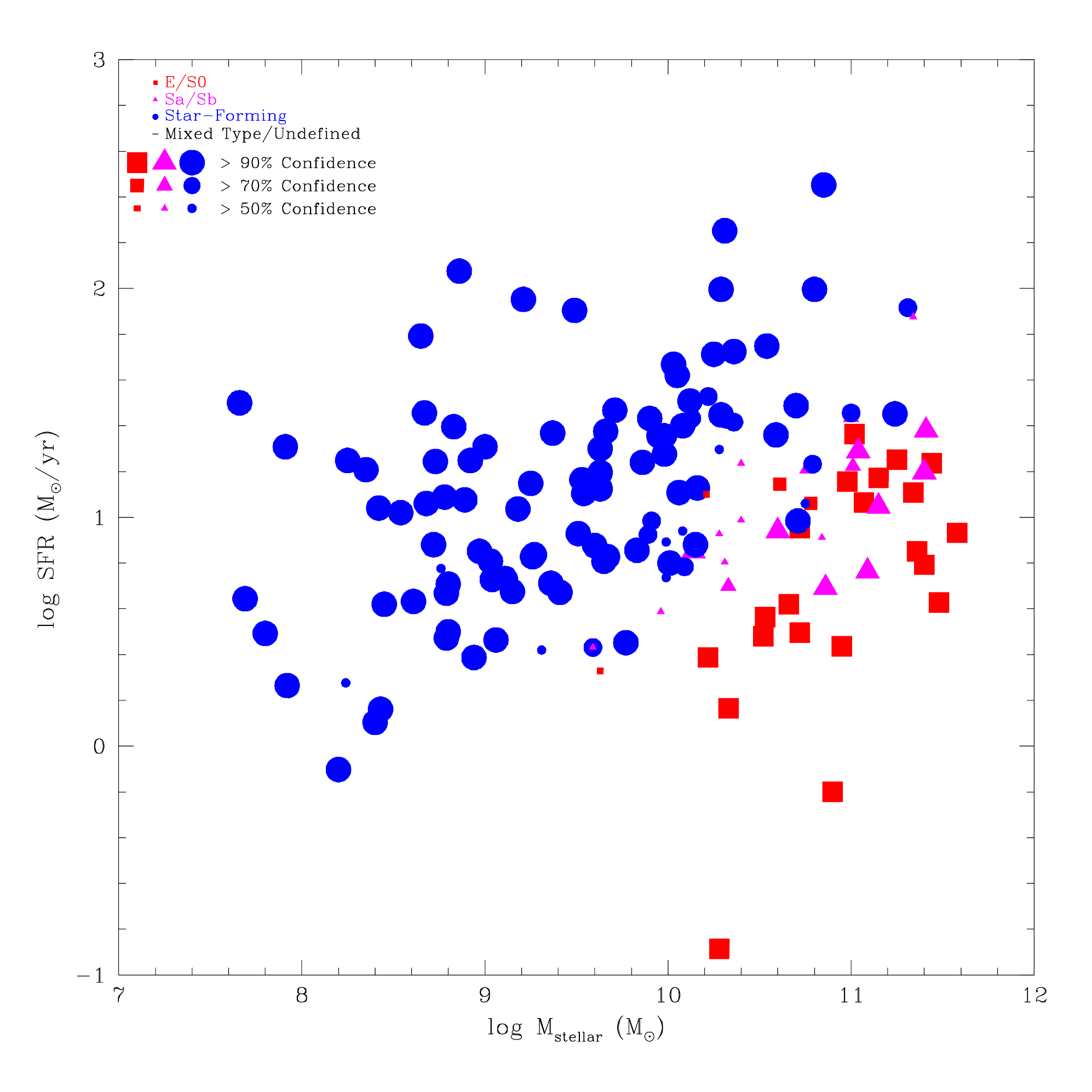}
\caption{Determined stellar masses versus SFR.  The galaxy type is based on our SED fit.  The size represents the corresponding confidence in the fit.  As expected, our sample consists of a large number of starburst galaxies.  What is somewhat surprising is the large number of elliptical galaxies with higher than expected SFRs.  However, this is seen in other studies, whereas elliptical galaxies at $z>0.5$ are frequently undergoing active mergers and do have residual star formation. }
\label{logMasslogSFR}
\end{figure}

\begin{figure}[htb]
\plotone{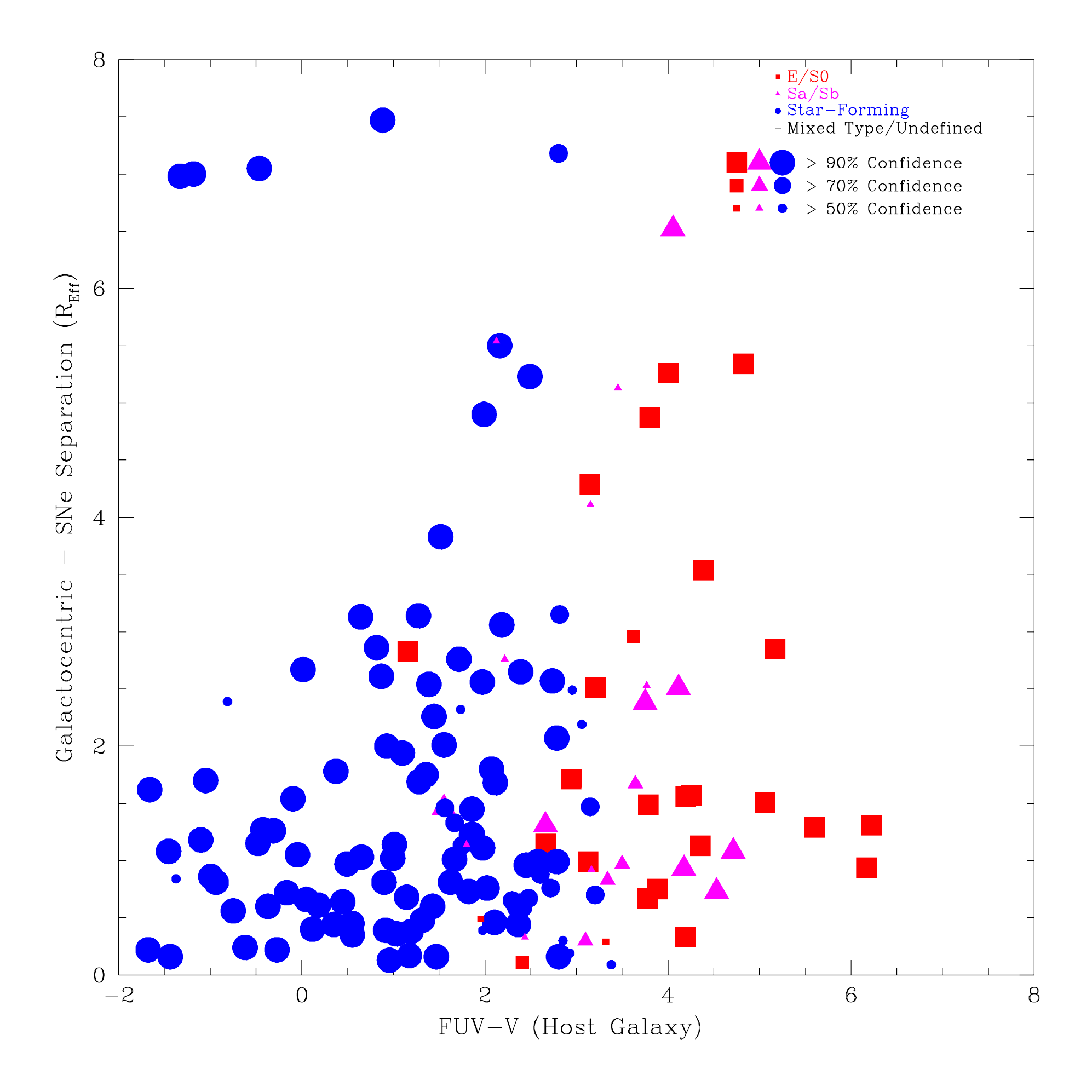}
\caption{Rest-frame $FUV-V$ color compared with the galactocentric separation of the SN normalized by the effective radius.  By normalizing it, we are able to compare the relative positions of SN in all galaxies types.  At $R/R_{Eff }> 2$, the SN are occurring environments with less stars and extinction, where we see $\approx 20\%$ of the SN occurring at these distances. }
\label{FUVVSNc}
\end{figure}

\begin{figure}[htb]
\plotone{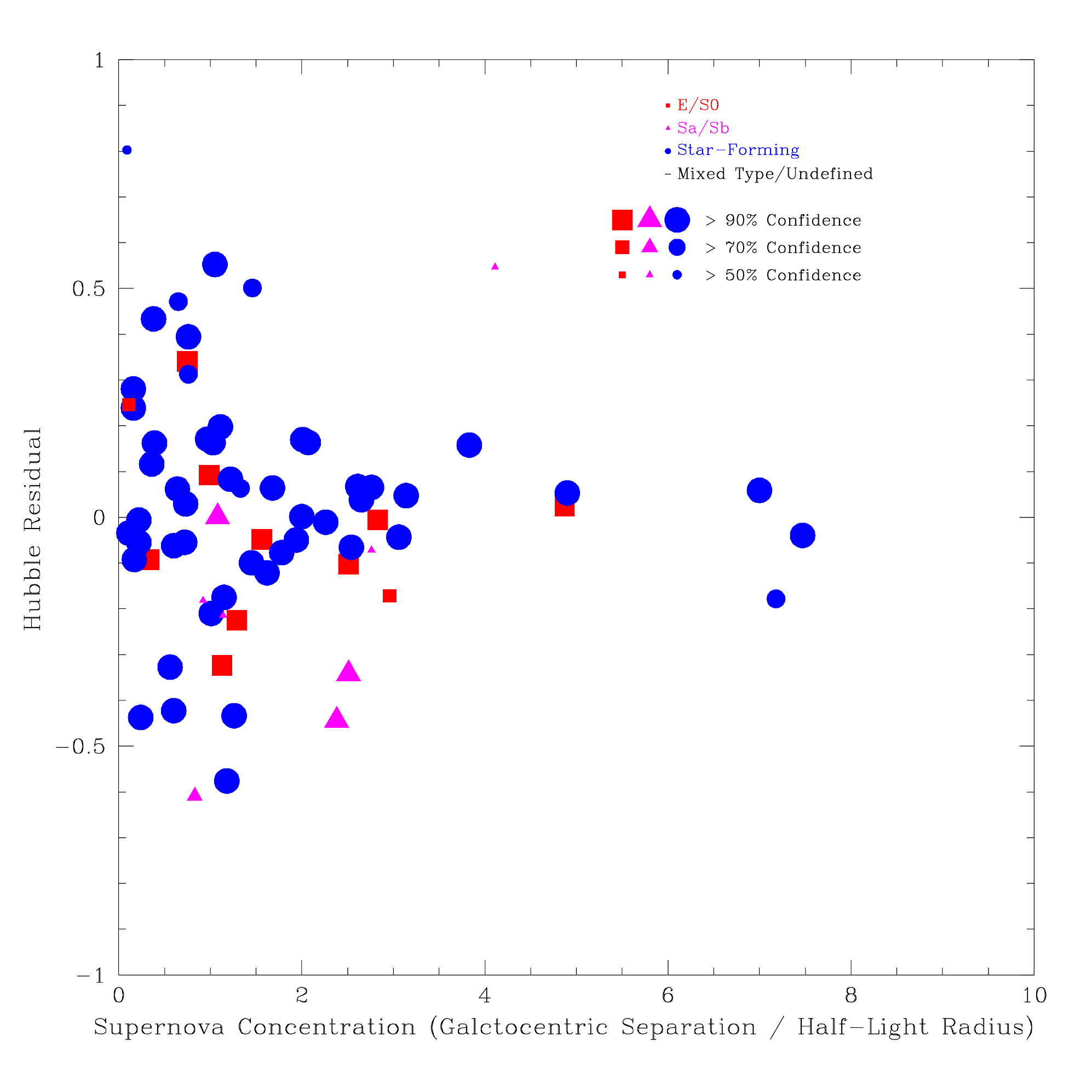}
\caption{Normalized supernova separation compared to the Hubble residual.  The residual is the offset from the best fit cosmology, where a positive value is a SN Ia that is fainter than expected, and a negative value is brighter.  We see that SN with $R/R_{Eff} > 2$ have better fits to the average cosmology.  At these large radii, we expect the average extinction of these objects to very low.}
\label{SNconcResid}
\end{figure}

\begin{figure}[htb]
\plotone{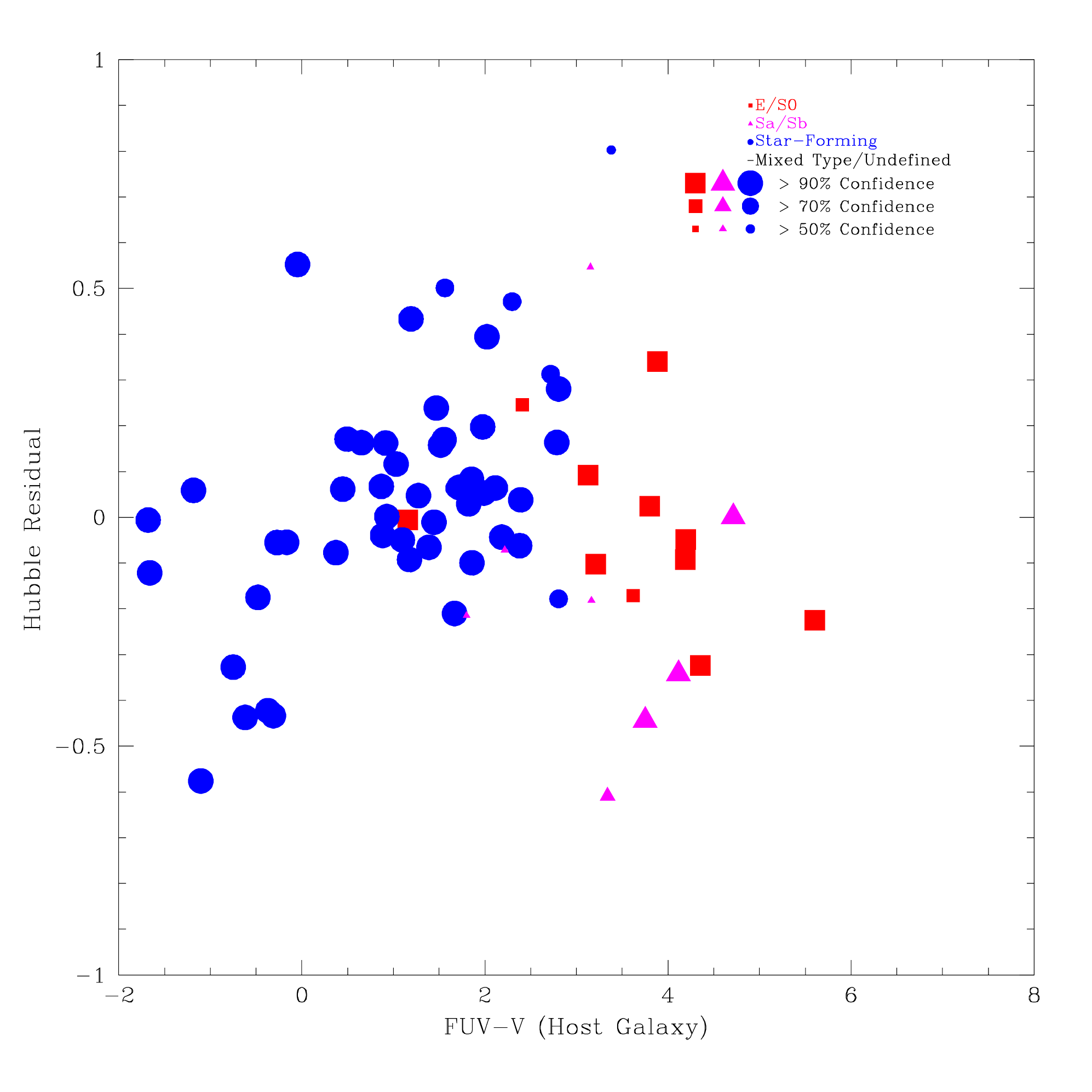}
\caption{The host galaxy $FUV-V$ color versus the SN Ia Hubble residual.  The residual is the offset from the best fit cosmology, where a positive value is a SN Ia that is fainter than expected, and a negative value is brighter.  We see a two component fit of SALT2 derived distances versus host Galaxy $FUV-V$ color.}
\label{FUVVResid}
\end{figure}

\section{UV Properties of Ia Host Galaxies}

\ The Equation of State SuperNova trace Cosmic Expansion (ESSENCE) project was a 6 year National Optical Astronomy Observatory (NOAO) survey program \cite[hereafter GM07]{Gajus07} which discovered 228 high redshift SNe Ia, $0.2<z<0.8$.  The primary goal of ESSENCE is to measure the dark energy equation of state parameter, $w = P/(\rho c^2)$, to better understand the nature of dark energy \citep{Kevin05,Davis07,MWV07}.  Using the MOSAIC II CCD on  the 4-m Blanco Telescope at the Cerro Tololo Inter-American Observatory (CTIO), ESSENCE took repeated images in broad-band Johnson - Cousin $R$ and $I$ filters in four fields, each covering 2 square degrees.  Supplemental imaging in Johnson-Cousin $B$ and $V$ and SDSS $z'$ for the purposes of host galaxy analysis was also undertaken.  Additionally, we have  \emph{Galaxy Evolution Explorer} (GalEx) data in the far-UV ($FUV$) and the near-UV ($NUV$) \citep{Martin05}.  The GalEx Medium Imaging Survey (MIS) overlaps with the majority of our ESSENCE fields, achieving a sensitivity $m_{AB}\approx 23$ \citep{Morrissey05}.

\ In the case of $BVRIz'$ images, we first stack the individual images in order to maximize our sensitivity.  Given the large quantity of $R$ and $I$ filters, we place criteria on both the full width half maximum(FWHM) (5 pixels or 1.35\arcsec) and the sky background levels.  For all other filters, no image restrictions are placed.  We then convolve individual images to a common PSF, and transform to a common coordinate system.   These transformed and convolved images are combined using two pixel rejection methods, one that uses an average sigma rejection and another that uses a minimum/maximum rejection.  Both methods give zero-points, consistent within $0.02$ magnitudes of each other.  For each SN, we identify all galaxies within a 10\arcsec($\approx 60$ Kpc) radius of the SN positions.  We identify the corresponding host galaxy by comparing the peak SN flux with the galaxy luminosity.  In some cases, a host galaxy could not be unambiguously identified, or was obscured by a foreground object.  Due to the large FWHM for GalEx, we require that the $FUV$ and $NUV$ have detections in both filters and positions within 2\arcsec of the optical positions.  This confirms the presence of the host in the UV and limits mis-identification. 

\ Photometric measurements are done using Source Extractor (SExtractor) \citep{SExtractor}, and we use the MAG\_AUTO photometry, which is based on Kron radius \citep{kron80}. \citet{GD05} showed that the Kron photometry in SExtractor is accurate to $90\% \pm 5\%$ of the total galaxy flux.  We perform relative photometry using catalogs from the ESSENCE project, limiting the calibration stars  to $16.5 < m_R < 21.0$, and fit the zero-point from this catalog for each galaxy.  We correct for Milky Way extinction using the \citet{Schlegel98} dust maps.  

\ Comparing galaxies at various distances is not trivial because the spectral energy distributions (SED) of the galaxies are highly varying and unknown while broad-band filters cover different parts of the spectrum at different redshifts.  We wrote a  k-correction procedure \citep{ESSENCEHosts} that converts observed frame broad-band photometry to any rest-frame magnitude system.  We make a simplifying assumption that the spectral energy distribution of any galaxy we observe can be described as a linear combination of locally observed galaxy types, ranging from ellipticals to starbursts. We use 11 galaxy SED templates, covering different morphological types  \citep{kinney96} and also a range of starburst galaxies with various extinction levels \citep{cksb94} obtained from \citet{bl03}. These templates cover a wavelength range of 100-1000nm, meaning we can transform observations between Johnson-Cousin \citep{mikeyb90}, SDSS, 2 Micron All Sky Survey (2MASS) \citealt{Cohen03,2MASS}, and Galaxy Evolution Explorer (GalEx) (\citealt{Martin05,Morrissey05}) photometric systems.. 

\ We cover the range of all galaxies using a series of linear combinations and define a $ \chi^2$ statistic based on the quality of fit to a template.   We allow a zero-point (ZP) offset due to the combined effects of intrinsic luminosity, distance, and template brightness, which we marginalize over.  This allows us to define a set of acceptable galaxy templates with varying fractions of each galaxy type, and the allowed range of zero-points, fit by our observer's frame photometry within $\Delta\chi^2 = 1$ .  We also allow upper limits, as we have non-detections in GalEx of some host galaxies.  Lastly, we also calculate the probability of a template based on a  goodness-of-fit test, providing us with a confidence of galaxy type.  

\ We also calculate physical parameters for our host galaxies.  Firstly we calculate the separation from the galactocentric position and the SN position using the respective positions.  We then calculate the galaxy scale length and size using the respective radii in SExtractor, FLUX\_RADIUS for our $50\%$ our effective radius, and the KRON\_RADIUS for our total galaxy size.  We normalize the galactocentric separation by the effective radius, enabling us to compare the position of objects with their host galaxies which accounts for the variation of host galaxy size.   

\ Our star formation rates (SFRs) are calculated using the empirical relation from \citet{Salim07}.   \citet{Salim07} used a large set of galaxies with GalEx UV and SDSS optical imaging to develop an empircal conversion between $L_{FUV}$ and SFR using \citet{BCmodels} models with a \citet{Chabrier03} initial mass function (IMF).  Using this conversion, we also calculate SFR upper limits for host galaxies with null detections in GalEx.  However, due to contamination from blue horizontal branch stars which emit in the $FUV$ filter,  we need to apply a nominal correction for this added flux \citep{Brown00}.  While the added flux is negligible compare to the emission from a starburst galaxy, it is necessary for an early type galaxy with little or no UV emission.  

\  Lastly, we determined stellar masses of our host galaxies using our $B-V$ color and the relations of \citet{BelldeJong} and their preferred model which uses \citet{BCmodels} with a $Z=0$ metallicity, Scaled Salpeter IMF and star formation bursts in the formation epoch.   We use determined $V$ band stellar mass, however  \citet{BelldeJong} normalize their relations so that all filters yield the same stellar mass so all filters should give the same value.  As these relations are best used for disk galaxies, we fit each galaxy using a  S\'ersic profile \citep{Sersic63}, where in we determine the S\'ersic index either as an exponential profile or the de Vaucouleurs profile which is applied for elliptical galaxies.  This fitting enables us to apply a nominal correction, making the application of the \citet{BelldeJong} relations relevant.  

\ In figure \ref{logMasslogSFR}, we compare our values of stellar mass and SFR.  It is interesting to note that some early type systems have relatively high SFRs.  This has not been previously seen in photometric studies as they did not use UV imaging as we have.  However, this effect has been seen in some elliptical galaxy spectra \citet{JG08}.  At $z>0.5$, the universal SFR density has already increased by a factor of 4, and it is more common to see star formation associated with early-type galaxies at such redshifts \citep{Rich05}.  UV photometry is one of the best indicators of SFR \citep{Kennicutt98,Adelberger00} and so we are able to obtain a accurate measurement of the true SFR of SN Ia host galaxies. 

\ Figure \ref{FUVVSNc} shows how are SN are distributed in their respective galaxies with respect to the type.  At separations $> 2 R_{Eff}$, a SN is occurring away from the bulk of its host galaxy, and consequently the bulk of stars and extinction.   Approximately $20\%$ of the SN Ia occur at these distances, suggesting a population of SN Ia occurring in lower extinction environments.  

\ Figure \ref{SNconcResid} plots this separation with respect to Hubble residual.  Here, a Hubble residual of 0 means that the SN distance was a perfect fit to the average cosmological values.  A positive residual means the distance was underestimated, and it is fainter than expected while a negative value is a SN that is brighter than expected.  Not surprising, the scatter for objects occurring near the center of their host galaxies ($R/R_{Eff} < 2$) have a larger scatter on the Hubble diagram \citep{Riess98}.  This probably is a consequence of more significant dust extinction experienced by SN Ia in these areas, as objects ($R/R_{Eff} > 2$) have less scatter where extinction is less likely.  This potentially provides a way to use SN Ia more effectively for cosmological measurements, circumventing some of the extinction problems. 

\ Lastly, we examine the properties of the cosmology fits as a function of the host galaxy properties., to see if we can see the stellar mass/metallicity effect as seen by others.  Figure \ref{FUVVResid} shows Hubble residuals as a function of $FUV-V$ color.   We see a two component trend, dependent on $FUV-V$, where both the bluest host galaxies (ones with high specific rates of star formation), and the reddest galaxies (old, red-dead galaxies) have light curve shape corrected distances which are systematically brighter than the average SN Ia.  This trend of brighter distances as a function of blue host galaxy $FUV-V$ colors been previously undetected again due to the lack of UV imaging.  \citet{Nomoto03} predicted that both metallicy and age are important ingredients in making a SN Ia.  The blue $FUV-V$ host galaxies suggest young stars, while the red $FUV-V$ host galaxy color correlates with both the mass, mean stellar age, and metallicity of the host.  With good UV data, these trends can be removed from SN cosmology fits. 

\section{Conclusions}

\ The UV portion of the spectrum contains highly useful pieces of information.  By looking at early time light curves in the ultraviolet, we do not see a signature of a red giant companion, as predicted by \citet{Kasen10}, casting doubt on the frequency of this channel as a progenitor of SN Ia.  Ultraviolet imaging of SN Ia host galaxies has yielded a wealth of information, including a two component residual in Hubble diagrams as a function of fost galaxy $FUV-V$ color.   The ultraviolet has also shown some early-type systems have some star-formation and it is probably not possible to assume that objects occurring in these systems are free of extinction.  What is possibly better will be to use the SN Ia's position within its host galaxy to help choose low extinction candidates.  With additional and improved UV data, such as that of the World Space Observatory - Ultraviolet (WSO-UV) \citep{WSOUV1,WSOUV2}, measurements of Dark Energy and our understanding of SN Ia will be greatly improved.

\bibliographystyle{../aa}
\bibliography{../citations.bib}

\end{document}